\documentclass[preprint,12pt]{aastex}
\newcommand{\kms}{km s$^{-1}$}
\shorttitle{Spurs in Spiral Galaxies}
\shortauthors{Shetty and Ostriker}
\begin{document}
\slugcomment{Accepted for publication in the Astrophysical Journal}
\title{Global Modeling of Spur Formation in Spiral Galaxies}
\author{Rahul Shetty and Eve C. Ostriker}
\affil{Department of Astronomy, University of Maryland, College Park, MD 20742-2421}
\email{shetty@astro.umd.edu, ostriker@astro.umd.edu}
\begin{abstract}
  
  We investigate the formation of substructure in spiral galaxies
  using global MHD simulations, including gas self-gravity.  Local
  modeling by Kim and Ostriker (2002) previously showed that
  self-gravity and magnetic fields cause rapid growth of overdensities
  in spiral arms; differential compression of gas flowing through the
  arms then results in formation of sheared structures in the
  interarms.  These sheared structures resemble features described as
  spurs or feathers in optical and IR observations of many spiral
  galaxies.  Global modeling extends previous local models by
  including the full effects of curvilinear coordinates, a realistic
  log-spiral perturbation, self-gravitational contribution from 5
  radial wavelengths of the spiral shock, and variation of density and
  epicyclic frequency with radius.  We show that with realistic Toomre
  $Q$ values, self-gravity and galactic differential rotation produce
  filamentary gaseous structures with kpc-scale separations,
  regardless of the strength -- or even presence -- of a stellar
  spiral potential.  However, a sufficiently strong spiral potential
  is required to produce ``true spurs'', consisting of interarm
  structures emerging from gas concentrations in the main spiral arms.
  In models where $Q$ is initially constant, filaments due to interarm
  self-gravity grow mainly in the outer regions, whereas true arm
  spurs grow only in the inner regions.  For models with $Q \propto
  R$, outer regions are intrinsically more stable so ``background''
  interarm filaments do not grow, but arm spurs can develop if the
  spiral potential is strong.  Unlike independently-growing
  ``background'' filaments, the orientation of arm spurs depends on
  galactic location.  Inside corotation, spurs emanate outward, on the
  convex side of the arm; outside corotation, spurs grow inward, on
  the concave side of the arm.  Based on orientation and the relation
  to arm clumps, it is possible to distinguish ``true spurs'' that
  originate as instabilities in the arms from independently growing
  ``background'' filaments.  We measure spur spacings of
  $\sim$3 - 5 times the Jeans length in the arm, and arm clump
  masses of $\approx10^7 M_\odot$.  Finally, we have also studied
  models without self-gravity, finding that magnetic fields suppress a
  purely hydrodynamic instability recently proposed by Wada \& Koda
  (2004) as a means of growing interarm spurs and feathers.  Our
  models also suggest that magnetic fields are important in preserving
  grand design spiral structure when gas in the arms fragments via
  self-gravity into GMCs.

\end{abstract}


\section{Introduction}

Observations of disk galaxies reveal that arm substructures are
prevalent in grand design spirals.  Some of the most prominent
secondary features are spurs and feathers, which are structures that
emanate from the primary spiral arm and are usually seen to sweep back
to trail the flow in the interarm region.  Historically, in the
observational literature the term ``spur'' has denoted stellar
features seen in optical emission (Elmegreen 1980), whereas the term
``feathering'' has been used to denote a series of extinction features
that overlies the bright portion of a stellar spiral arm (Lynds 1970).
Elmegreen (1980) concluded that spurs are long-lived features, based
on observations in multiple bands, including the I band, which
generally traces the older stellar component.  Recent high-resolution
observations, such as the Hubble Heritage image of M51 (Scoville \&
Rector 2001) and several galaxies in the Spitzer SINGS sample
(Kennicutt et al.\ 2003), have revealed examples of these features in
extraordinary detail.  An archival study of HST images has established
that spurs and feathers are in fact ubiquitous in galaxies with
well-defined spiral arms and that single continuous structures
evidence evolution from primarily gaseous to primarily stellar
composition (La Vigne et al.\ 2006).  Taken together, these observations
indicate that feathers and spurs are an essential aspect of spiral
structure that should be accounted for by theoretical modeling of disk
galaxies.

Many studies of spiral structure, both theoretical and observational,
have applied the hypothesis of a Quasi Stationary Spiral Structure
(QSSS) (Lin \& Shu 1964).  Under the QSSS framework, the general shape
of the spiral pattern is assumed to remain steady for many galactic
revolutions.  The stellar spiral arms themselves arise as
self-consistent density waves (or modes).  Roberts (1969) demonstrated
that shocks can develop in the gaseous component as it responds to an
``external'' spiral potential arising from the stellar disk, and
predicted values of the gas velocity both upstream and downstream from
the shocks.  Such velocity profiles have been observed for many
galaxies, such as M51 (e.g. Rand 1993, Aalto et al.\ 1999, Shetty et
al.\ 2006), and M100 (e.g. Rand 1995).  M81 has also been studied
extensively in support of density wave theory (e.g. Visser 1980 and
Lowe et al.\ 1994).

A number of theories consistent with the QSSS concept have been
proposed to explain substructure in spiral galaxies.  Shu, Milione, \&
Roberts (1973) suggest that ultra-harmonic resonances between the
motion of the primary spiral pattern and the background gas flow can
be responsible for the secondary features in spiral galaxies.
Chakrabarti, Laughlin, \& Shu (2003, hereafter CLS) performed
hydrodynamic simulations of self-gravitating gaseous disks to study
the role of ultraharmonic resonances.  They showed that the spiral arm
bifurcates and strong branches, which in observations are large scale
dust lanes that are similar in angular extent to the main arms, occur
near resonant radii (see also Artymowicz \& Lubow 1992).

The origin and nature of observed smaller-scale feathers and spurs,
however, has not yet been firmly established.  Balbus (1988)
attributed these features to growth of gravitational instabilities in
the gas component in preferred directions.  Kim \& Ostriker (2002),
hereafter KO, performed numerical simulations focused on a local patch
of a gaseous spiral arm in a galactic disk, and showed that the growth
of prominent, nonlinear spurs\footnote{We adopt the term ``spur'',
  following KO, to describe interarm gas features in hydrodynamic and
  MHD models.} can occur due to the mutual contributions of
self-gravity and magnetic fields, via the so-called ``Magneto-Jeans
Instability.''  Within the arm, the radial gradient in angular
velocity is reversed, so that spurs in the models of KO are initially
locally leading.  Well into the interarm regions, background galactic
shear causes the spurs to become trailing features.  This
characteristic shape is evident in the Hubble Heritage image of M51
(Scoville \& Rector 2001).  On the other hand, Wada \& Koda (2004),
hereafter WK, suggest that the growth of spurs results from purely
hydrodynamic effects.  In their two-dimensional models (excluding
magnetic fields and self-gravity), the spiral shocks become unstable;
this instability causes the growth of clumps and subsequently leading
interarm features, which they refer to as spurs.  They suggest the
mechanism responsible for the growth of these spurs is the
Kelvin-Helmholtz instability.  Using SPH simulations, Dobbs \& Bonnell
(2006) also investigate the non-self-gravitating case, and suggest
that feather and spur formation requires gas temperatures $<$ 1000K.
Contemporary with the current work, Kim \& Ostriker (2006) extended
their 2D local self-gravitating models to 3D; that work also
investigates the effects of vertical structure on hydrodynamic
Kelvin-Helmholtz modes.

Here, we model isothermal gaseous disks under the influence of both
magnetic and self-gravitational effects.  We extend the local
simulations of KO into the global regime, which allows us to study the
growth of spurs using more realistic models.  In particular, whereas
in KO the unperturbed disk had uniform density, a linear shear
profile, and did not treat curvature effects, the present models relax
all of these idealizations.  Our global models, which extend over more
than an order of magnitude in radius, allow arbitrary density profiles
and rotation curves, and solve the full equations of
magnetohydrodynamics (MHD) in cylindrical symmetry.  To assess the
effect of self-gravity, we first consider just a disk of rotating gas,
varying initial conditions such as the magnetic field strength and
Toomre stability parameter.  We then apply an external spiral
potential, which reorganizes the gas to form a spiral pattern.  In
models with and without a spiral potential, we follow the evolution of
the gas far into the nonlinear domain as self-gravity takes hold,
investigating the properties of the interarm features and clumps that
arise.

This paper is organized as follows: In $\S$\ref{method} we present the
relevant equations of MHD and gravity, and describe the models,
parameters, and numerical algorithms we use to simulate disk galaxies
with spiral arms.  In the following section ($\S$3), we consider
models without including self-gravity, both with and without magnetic
fields.  Next, in $\S$\ref{sgsec}, we present the models including
self-gravity.  We show how self-gravity causes the growth of
condensations in the gas, and how this effect is crucial for the
growth of interarm structures in disks with an external spiral
potential.  In $\S$\ref{disc} we analyze and discuss various aspects
of our results, including: in $\S$\ref{asp}, we explore the issue of
distinguishing whether observed interarm structures are true arm spurs
or independent background features with a superposed large-scale
spiral structure; in $\S$\ref{clumps} we quantify masses and spacings
of clumps and spurs; and in $\S$\ref{gccomp} and $\S$\ref{Qeff} we
discuss the issue of gas/star arm offsets and disk thickness,
respectively.  We conclude in $\S$\ref{summ} with a summary.  Finally,
in the Appendix, we give a detailed description of our algorithms to
compute self-gravity, as well as tests comparing the results of
different numerical approaches.

\section{Modeling Methods \label{method}}
\subsection{Basic Equations}
Our simulations involve the gaseous response to an external spiral
potential, including effects of self-gravity and magnetic fields, in a
two-dimensional galactic disk model.  The gas is initially in pure
circular motion around the galactic center.  We adopt a flat rotation
curve, i.e.\ a constant azimuthal velocity $v_c$.  A non-axisymmetric
variation in the stellar component is responsible for an external
spiral perturbation, and is modeled as a rigidly rotating potential
with a pattern speed $\Omega_p$.  We investigate the formation and
evolution of arms, spurs, clumps, and other features by integrating
the MHD equations in a polar $(R,\phi)$ coordinate system.

The relevant equations of MHD and gas self-gravity are
\begin{equation}
  \frac{\partial \Sigma}{\partial t} + \nabla \cdot (\Sigma {\bf v} ) =
  0,
\label{cont}
\end{equation}
\begin{equation}
  \frac{\partial {\bf v}}{\partial t} + {\bf v}\cdot\nabla {\bf v} +
  \frac{1}{\Sigma}\nabla p = \frac{2 H}{4 \pi \Sigma} (\nabla \times {\bf
    B})\times {\bf B} - \nabla(\Phi_{ext} + \Phi), 
\label{force}\
\end{equation}
\begin{equation}
  \frac{\partial {\bf B}}{\partial t} = \nabla \times ({\bf v} \times
  {\bf B}), 
\label{Faraday}
\end{equation}
\begin{equation}
  \nabla^2\Phi = 4\pi G \delta(z) \Sigma.
\label{Poisson}
\end{equation}
Here, $\Sigma$ is the gas surface density, and ${\bf v}$, $p$, and
${\bf B}$ are the vertically averaged velocity, vertically integrated
pressure, and vertically averaged magnetic field, respectively.  The
semi-thickness of the disk is $H$, such that $\Sigma / 2H$ is the
mid-plane density $\rho_0$.  For our models, we assume an isothermal
equation of state, so that $p = c_s^2\Sigma$, where $c_s$ is the sound
speed.  The terms $\Phi_{ext}$ and $\Phi$, respectively, represent the
external spiral potential and the gaseous self-gravitational
potential.  The external spiral potential $\Phi_{ext}$ is specified at
time $t$, in the inertial frame, by
\begin{equation}
\Phi_{ext}(R,\phi;t)=\Phi_{ext,0}\cos[m \phi - \phi_0(R) - m\Omega_pt]
\label{spiralpot}
\end{equation}
where $m$, $\phi_0(R)$, and $\Omega_p$ are the number of arms,
reference phase angle, and spiral pattern speed, respectively.  We
assume the spiral arms have a constant pitch angle $i$, implying a
logarithmic spiral, so that
\begin{equation}
\phi_0(R) = -\frac{m}{\tan\,i}\ln(R) + constant.
\label{refang}
\end{equation}
Since we simulate disks in an inertial frame of reference, explicit
corrections to equation (\ref{force}) for Coriolis and centrifugal forces are not required.  

\subsection{Model Parameters}
For the simulations presented in this paper, we use a constant sound
speed $c_s$, which we set to 7 \kms~ for scaling our solutions.  We
use a constant rotational velocity $v_c$, which is set to 210 \kms .
Because our models are isothermal, in fact our results would hold for
any model with the same ratio $v_c / c_s$ = 30.  The code length unit
is $L_0$, which for convenient scaling of our solutions we set to 1
kpc.  With $c_s$ = 7 km s$^{-1}$, this implies a time unit for scaling
of $t_0 = L_0 / c_s = 1.4 \times 10^8$ years.  This time corresponds
to one orbit $t_{orb} = 2 \pi/\Omega_0$ at a fiducial radius $R_0$
which is given by $R_0 = L_0 v_c / 2\pi c_s$.  With $L_0$ = 1 kpc, the
value of $R_0$ is 4.77 kpc.  Our results can be rescaled to other
values of $R_0$ and $L_0$ with the same ratio.

We explore a range of values for the parameters required to specify
the spiral perturbation.  The amplitude of the potential perturbation
$\Phi_{ext,0}$ in equation (\ref{spiralpot}) is characterized by the
ratio $F$ of the maximum radial perturbation force to the radial force
from the background axisymmetric potential (responsible for $v_c$),
i.e.
\begin{equation}
F\equiv\frac{\Phi_{ext,0}m}{v_c^2\tan i}.
\label{eps}
\end{equation}
We model spiral disks with external potential strengths $F$ = 10\%,
3\%, and 1\%.  We apply the spiral perturbation gradually, increasing
from zero and settling to the maximum level $F$ at time $t = t_{orb}$.
Since it has proved to be difficult to locate corotation from
observations, we also explore a range in $\Omega_p / \Omega_0$ = 0.19
to 0.96.  For our fiducial values of $L_0$ and $c_s$, $\Omega_p$
ranges from $\sim$8 to 42 \kms $\,$kpc$^{-1}$, corresponding to
corotation radii of 25 to 5 kpc for a circular velocity of 210 \kms.
The pitch angle $i$ in most of our simulations is $10\degr$.  We will
show that changing the pitch angle does not strongly affect the
formation and properties of substructures.  For all our models, we
have an $m$ = 2 (two-armed) pattern.

The initial surface density $\Sigma_0$ at the fiducial radius
$R_0$, along with the constant circular velocity $v_c$, determines
the value of the Toomre stability parameter at $R_0$,
\begin{equation}
Q_0 \equiv \frac{\kappa_0 c_s}{\pi G \Sigma_0},
\label{Q}
\end{equation}
where $\kappa_0$ is the initial epicyclic frequency; for a
constant circular velocity, $\kappa_0 = \sqrt2 \Omega_0 = \sqrt2
v_c/R_0$.  Thus, the initial background surface density at $R_0$ is
\begin{equation}
\Sigma_0 = \frac{32}{Q_0}\, M_\odot\, {\rm pc}^{-2}
\left(\frac{c_s}{7\,{{\rm km\, s}^{-1}}}\right)\left( \frac{\kappa_0}{62\,
    {\rm km\, s}^{-1}\, {\rm kpc}^{-1}}\right).
\label{inits}
\end{equation}
As described below, we explore a couple of initial density
distributions, including a $\Sigma \propto R^{-1}$ density distribution
for which $Q$ is initially constant for the whole disk.

We characterize the initial magnetic field strength by the ratio $\beta$
of the midplane gas pressure to the midplane magnetic field pressure
\begin{equation}
\beta = \frac{P_{gas}}{P_B} = \frac{8 \pi c_s^2 \Sigma}{2HB^2}. 
\label{beta}
\end{equation}
The initial magnetic field lines ${\bf B} = B\hat\phi$ lie in the
plane of the disk, and are directed in the $\hat\phi$ direction only.
From equation (\ref{beta}) the value of the magnetic field (with
$c_s$, $\beta$, and $H$ constant) varies $\propto \Sigma^{1/2}$.
Taking $H$ = 100 pc and $Q_0$ = 2, the value of $B$ at the fiducial
radius is
\begin{equation}
B_0 = \left(\frac{4c_s^3\kappa_0}{Q_0HG\beta}\right)^{1/2} = \frac{8.2}{\sqrt{\beta}}\, \mu G. 
\label{B0beta}
\end{equation}

\subsection{Numerical Methods \label{nummet} }
We follow the evolution of the gaseous disk by integrating equations
(\ref{cont}) - (\ref{Poisson}) using a cylindrical polar version of
the ZEUS code.  ZEUS (Stone \& Norman, 1992a, 1992b) is a
time-explicit, operator-split, finite difference method for solving
the equations of MHD on a staggered mesh.  ZEUS employs ``constrained
transport" to ensure that $\nabla\cdot {\bf{B}} = 0$, and the ``method
of characteristics" for accurate propagation of Alfv\'{e}nic
disturbances.  The hydrodynamic/MHD portion of our cylindrical-polar
code has been verified using a standard suite of test problems.  These
include advection tests, shocks aligned and not aligned with the
coordinates, a magnetized rotating wind (Stone \& Norman 1992b), and a
rotating equilibrium disk with both magnetic and pressure gradients.

The $(R,\phi)$ staggered mesh for our version of the code has a
constant logarithmic increment in the radial dimension, i.e.\ $R_{i+1}
= (1 + \delta)R_i$, for some $\delta > 0$.  We consider only the
perturbed gas density (by subtracting off the initial density) in
determining the self-gravitational potential at each timestep.  The
contribution of the initial axisymmetric gas disk to the total
potential is assumed to be included in the axisymmetric potential
responsible for the constant circular velocity $v_c$.  To compute the
self-gravitational potential $\Phi$, we use one of two methods
described in the Appendix.  One method uses a combination of a Fourier
transform in the azimuthal direction and a Green's function in the
radial direction, while the other method uses Fourier transforms in
both directions on an expanded, zero-padded grid.  For both methods,
we allow for finite disk thickness using a softening parameter $H$.
Except as noted, we adopt $H$ = 100 pc.  We give detailed descriptions
of our methods, tests, and comparisons in the Appendix.  Fast Fourier
transforms are performed using the free software FFTW (Frigo \&
Johnson 2005).

The standard computational domain for our models has 512 radial and
1024 azimuthal zones, covering a radius range of 1 - 15 kpc, and an
azimuthal range of 0 - $\pi$ radians.  With this resolution, the Jeans
length ($\lambda_J = c_s^2/G\Sigma$) from our initial surface density
distributions are well resolved at all radii, satisfying the Truelove
criterion (Truelove et al.\ 1997).  The radial range allows $\approx$ 5
radial wavelengths of the spiral pattern.  Given the extended range in
the radial dimension, we implement outflow boundary conditions, since
loss of gas at the boundary will not affect the majority of the disk.
We also taper the spiral potential near the boundaries, which helps
minimize loss of matter near the edges.  In the azimuthal direction,
we use periodic boundary conditions.  Though the azimuthal range is
only half of a complete disk, the gravitational potential includes the
contribution from the other half that is not explicitly simulated (see
Appendix).

\section{Simulating Spiral Galaxies Without Gas Self-Gravity\label{nsg}}
We first investigate the flow of gas in a spiral potential without
including the gaseous self-gravitational potential.  These preliminary
simulations will indicate whether, for a given parameter set and
numerical resolution, long lasting spiral patterns can be sustained.
We will also investigate the effect of magnetic fields on the
resulting flow and spiral morphology.  As these models are similar to
the hydrodynamic simulations of WK, we are able to investigate the
``wiggle instability'' that they propose, and to assess how magnetic
fields affect this process.  Table \ref{models} shows the relevant
parameters used for each model.  Column (1) labels each model.  Column
(2) lists $\beta$, which characterizes the magnetic field strength
(see eq. [\ref{beta}]).  The external potential strength $F$ (from
eq. [\ref{eps}]) is listed in column (3).  Column (4) gives the pattern
speed $\Omega_p$ (used in eq. [\ref{spiralpot}]).  The pitch angle
$i$ is listed in column (5).  We note that though the computational
domain only simulates half the disk, with periodic azimuthal boundary
conditions, we replicate the simulated half in presenting snapshots of
the models.

\subsection{Pure Hydrodynamic Models}

We begin by considering simple cases where the rotating gas in a disk
only responds to an external spiral potential, without including
magnetic fields or gas self-gravity.  Model HD1 has a weak ($F$ = 3\%)
and slowly rotating external spiral potential, as well as a small
pitch angle (10\degr).  Figure \ref{HD1pl} shows density snapshots of
model HD1.  At $t/t_{orb}$ = 1, when the external potential reaches
its maximum amplitude, the spiral arms are weak but distinct.  Figures
\ref{HD1pl}(b) and \ref{HD1pl}(c) shows that 2 orbits after the
potential is fully applied, the spiral arms are still distinct and
rather regular.  At $t/t_{orb}$ = 3 , in the inner regions, shown
in Figure \ref{HD1pl}(d), the spiral arms are not as distinct as the
rest of the galaxy.  The main arms grow weaker, and leading spiral
like features grow between the main (trailing) arms.  Nevertheless, a
global spiral pattern persists throughout the galaxy, indicating that
a weak perturbing potential can sustain a global, long lasting, spiral
pattern.

Figure \ref{HD2pl} shows density snapshots of model HD2, with the same
scale shown in Figure \ref{HD1pl} for model HD1.  As expected, since
$F$ is increased to 10 \%, the spiral arms are much stronger.  The
global pattern persists for many orbits, but the arms are clearly more
dynamic.  As early as 1 orbit after the potential is fully applied (at
$t/t_{orb} = 2$), the spiral arms at $\sim$7.5 kpc, indicated by the
arrow in Figure \ref{HD2pl}(b), bifurcates.  This region is near the
Inner Lindblad Resonance (ILR).  The bifurcation causes the arm at
$\sim$7 kpc to lose matter, and it thus becomes weaker than the arms
located farther inward.  Further, the arm at $\sim$9.3 kpc has a much
different pitch angle from the arms at different locations.  After an
additional orbit, the bifurcated part of the inner arm has moved
radially outward and connected with the outer arms.  In the meantime,
the arms that lost matter during bifurcation regain strength and
attain similar surface density to the arms in the inner regions.

Figure \ref{HD2pl}(d) shows the central regions of model HD2 at
$t/t_{orb}$ = 3.  Here, unlike in the case with a weaker potential
(Fig.  \ref{HD1pl}(d)), the arms remain continuous and distinct.
However, there are also prominent interarm filamentary features, some
even connecting two adjacent arm segments.  Such features can be seen
to develop as early as $t/t_{orb}$ = 2, in Figure \ref{HD2pl}(b).  WK
found similar features, which they identified as spurs/fins, in their
hydrodynamic models (the detailed morphology differs because they use
a different rotation curve).  They attribute the formation of their
spurs/fins to the Kelvin-Helmholtz instability.  In our models, these
features only grow in the innermost regions; we shall show in the
following section that magnetic fields prevent their
formation.\footnote{Recent models by Kim \& Ostriker (2006) have also
  shown that the instability identified by WK is suppressed by
  three-dimensional effects even in unmagnetized models.}  We shall
further show that it is the combination of magnetic fields {\it and}
self-gravity that results in spurs forming everywhere in a disk, not
just in the innermost regions.

Another difference between models HD1 and HD2 is the
relative location of the gas density peaks of the arms.  The gaseous
arms in model HD1 ($F$ = 3\%) form farther downstream than in model
HD2 ($F$ = 10\%).  We discuss the offset between the dust lanes and
the spiral potential minimum in $\S$\ref{summ}, as well as compare
with the results from the recent study by Gittins and Clarke (2004).

Changing the pattern speed of the spiral potential does not
dramatically alter the resulting spiral structure.  Figure
\ref{HD3_4pl}(a) shows a snapshot of model HD3, one orbit after the
external spiral potential with $F=10\%$ is fully applied.  Here, the
corotation radius is at 5 kpc, instead of 25 kpc.  The spiral
structure is similar to that shown in Figure \ref{HD2pl}(b), but the
arms are not as dynamic, and the bifurcation region is shifted inward,
as expected if this phenomenon is indeed due to a resonance.  For all
our models the shock locus transitions from the concave to the convex
side of the gaseous arm at or near corotation.  Inside this radius,
the shock front is located on the concave side of the gaseous spiral
arm.  Farther out in the disk, the shock front moves to the outer,
convex side of the arm.

Figure \ref{profrel} shows the density and velocity profiles relative
to the external spiral potential in two regions inside and outside
corotation for model HD3. For both regions, the gas peaks occur
downstream from the minimum of $\Phi_{ext}$.  Inside corotation, the
gas shocks after the gas passes through the spiral potential.  Outside
corotation, the spiral pattern passes through the gas, leaving the
shocked gas behind.  The shock front itself is always upstream from
the density peak.  Thus, inside corotation the shock occurs on the
inside face of the spiral arm, while outside corotation the shock
forms on the outer face of the arm.  The density and velocity profiles
in Figure \ref{profrel} are quite similar to those obtained using
local models (e.g. Figs.  2, 3 of KO).  We note that in regions closer
to corotation, the gas peaks lie near the minimum in $\Phi_{ext}$, and
shocks cannot be clearly distinguished.

The secondary density hump in the profile inside corotation occurs
near the 4:1 ultraharmonic resonance, where $\Omega_p - \Omega =
-\kappa /4$ (if pressure effects are ignored).  CLS also identified
similar secondary features in their global hydrodynamic models, which
they denote as ``branches.''  They also find such branches near
locations of the 6:1 ultraharmonic resonance.  Qualitatively, the
formation and subsequent evolution of the branch features in our
models are similar to those shown in CLS.

Models with a larger pitch angle show some differences from those with
more tightly wrapped arms.  Figure \ref{HD3_4pl}(b) is a snapshot of
model HD4, again 1 orbit after the full spiral potential is applied.
Similar to the results of WK, this model shows that loosely wound
spiral arms are much more unstable than tightly wound arms, because
the shock is stronger.  The bifurcation is clearly evident, and
results in replenishment of the depleted arms in the outer regions
after an additional orbit.  The interarm sheared filamentary features
in the inner region of the galaxy are more pronounced than in the
corresponding model with $i$=10$\degr$, shown in Figure
\ref{HD2pl}(b).

\subsection{Magnetohydrodynamic Models}
Before including self-gravity, we test the effect of magnetic fields
on non-self-gravitating disks with a spiral potential.  Figure
\ref{MHDcomp} compares snapshots at $t/t_{orb} = 2.0$, of a model
without magnetic fields, HD2, to one with magnetic fields, MHD1.
Clearly, the interarm features described in the previous section no
longer appear in the magnetized case.  Thus, equipartition-strength
magnetic fields are able to suppress the ``wiggle instability''
identified by WK in large $F$ simulations (for small enough $F$, as
seen in model HD1, there is stability even in unmagnetized models).

\section{Models Including Gas Self-Gravity \label{sgsec}}

To include self-gravity in our simulations, we must introduce an
additional parameter, which we choose to be the Toomre parameter $Q_0$
evaluated at $R_0$, given in equation (\ref{Q}).  Table \ref{sgmodels}
shows the input parameters for models including self-gravity.  The
first five columns are the same as those in Table \ref{models}, and
column (6) gives the value of $Q_0$.  As shown in equation (\ref{Q}),
$Q \propto \kappa / \Sigma$.

\subsection{Disk Stability Tests for Constant $Q$ Models}

For disks with constant circular velocities, the epicyclic frequency
$\kappa \propto R^{-1}$; thus if the initial density distribution
$\Sigma \propto R^{-1}$, $Q$ will be constant for the whole disk.  We
first consider models in which the initial surface density profiles
are indeed inversely proportional to the galactocentric radius.  Such
a distribution is consistent with many surface density profiles shown
in Regan et al.\ (2001) and Wong \& Blitz (2002).

To test the inherent stability of disks with constant $Q$, we consider
cases with self-gravity, but with no external potential, models
SHDne1, SHDne2, and SMHDne, shown in Figure \ref{nosp}.  Since the
initial density has random white-noise 0.1\% perturbations, the
over-dense regions can grow due to self-gravity.  As these regions
grow, they also become stretched azimuthally due to the background
shear.  The runaway growth of the over-dense regions eventually causes
neighboring regions to have extremely large velocities, such that the
Courant condition would require an extremely small timestep; we
therefore halt the simulation.  Model SHDne1 has $Q_0$=1, and becomes
unstable very rapidly (within one orbit at $R_0$), as shown in Figure
\ref{nosp}(a).  Figure \ref{nosp}(b) shows model SHDne2, with $Q_0$ =
2, at the same time as the $Q_0$ = 1 model in Figure \ref{nosp}(a).
Since the $Q_0$ = 2 model is more stable, enough time has not yet
elapsed for the over-dense regions to dominate.

The addition of magnetic fields, as shown in Figure \ref{nosp}(c)
from model SMHDne with $Q_0$ = 1, does not affect the growth of
filaments significantly.  The subtle difference is that the magnetic
fields slightly slow the growth of over-dense regions.  As a result,
models including magnetic fields evolve longer before the flow
velocities in some zones becomes extreme.  Figure \ref{nosp}(d) is the
last snapshot of model SHDne2 at $t/t_{orb}$ = 1.5 orbits.  Here, the
outer regions have evolved to the point that the structure is similar
to that in Figure \ref{nosp}(a).  However, it is clear that the radius
of the stable inner region in the $Q_0$ = 2 model is larger than that
of the $Q_0$ = 1 model (SHDne1).  As expected, increasing the value of
$Q_0$ increases the area of stability in the inner regions, and
requires more time for the instability in the outer regions to grow.

The stability tests show that filament-like structures will grow in a
shearing disk with sufficient gas surface density, regardless of the
presence of magnetic fields.  The Toomre stability parameter governs
which regions are prone to gravitational instabilities.  As $Q_0$
increases, the outer disk becomes more stable, and more time is
required for growth of instabilities.  In the models presented thus
far, we have considered disks for which the initial surface densities
vary as $R^{-1}$, so that $Q$ is initially constant everywhere in the
disk.  The reason that the outer disk becomes unstable even when the
inner disk does not is that the disk thickness $H$ is constant throughout
the disk.  The finite thickness stabilizes the inner disk more than
the outer disk, because the ratios of $H/\lambda_T$ and $H/\lambda_J$
vary as $R^{-1}$ for constant $Q$ models, where the Toomre wavelength
$\lambda_T = 4 \pi^2 G \Sigma / \kappa^2$ and the Jeans wavelength
$\lambda_J = c_s^2 / G \Sigma$.  We discuss this effect in
$\S$\ref{Qeff}.

\subsection{Disk Stability Tests for $Q \propto R$ Models} 

The models we have presented so far have an initial surface density
distribution proportional to $R^{-1}$, yielding a constant value of the
Toomre parameter with $R$.  However, the surface density
distributions shown by Wong \& Blitz (2002) are in many cases
approximately consistent with a surface density distribution
proportional to $R^{-2}$.  Further, in the observational analysis of
Martin \& Kennicutt (2001), $Q$ varies with radius for many galaxies.
A variety of radial distributions are evident, some close to $R^{-1}$
and others to $R^{-2}$.  We thus consider models similar to those
presented thus far, but with initial $R^{-2 }$ surface density
distributions, such that $Q \propto R$.  The labels of such models
will follow the convention of those already presented, but with the
addition of a prime ($\prime$) sign.

Figure \ref{Qvartest} shows snapshots of disk models without an
external potential, SHDne1$^\prime$ and SHDne2$^\prime$, at $t/t_{orb}
= 1.0$.  For SHDne1$^\prime$, the value of $Q$ ranges from 0.21 at the
inner boundary to 3.15 at the outer boundary.  The respective values
are twice as large in model SHDne2$^\prime$.  When compared with
Figure \ref{nosp}, it is clear that these models are much more stable.
Only the innermost region in SHDne1$^\prime$ shows more instability
than SHDne1, the corresponding disk with $\Sigma$ initially $\propto
R^{-1}$ (Fig. \ref{nosp}(a)).  As expected, since $Q$ increases with
$R$, the outer regions of the disk are more stable, and thus less
susceptible to gravitational instabilities.  In fact, there has been
very little growth of perturbations in model SHDne2$^\prime$ ($Q_0$ =
2) at 1 orbit (Fig. \ref{Qvartest}(b)).  However, given enough time,
the instabilities eventually begin to grow in this disk, and will
appear similar to the snapshot in Figure \ref{Qvartest}(a).

\subsection{Spiral Models with Constant $Q$}

To investigate the interaction between gaseous self-gravity and the
global spiral structure, we focus our presentation on six spiral
models with parameters shown in Table \ref{sgmodels}.  In addition to
these models, we have performed additional simulations with a wide
range of values and combinations of the chosen parameters, with
similar characteristic results.  For our fiducial model, $\beta=1$,
$\Omega_p$ = 8.4 (corresponding to a corotation radius of 25 kpc,
which is outside the edge of the disk), $i$ = 10\degr, and $Q_0$ = 2.
The external potential strength $F$ will be indicated, as will cases
where the other parameters differ from the fiducial one.

Figure \ref{SMHD1_2pl} shows the fiducial model with $F$ = 3\% and $F$
= 10\%, SMHD1 and SMHD2, respectively, at $t/t_{orb}$ = 1.0 and 1.125.
The most striking aspect of the snapshots, besides the spiral arms,
are the interarm features.  These features differ significantly
between the two models, with differences enhanced at the later times.
This is shown for the central region in detail in Figure
\ref{SMHD1_2inpl}.  With a weak external potential (SMHD1), the
interarm features are strong as far inwards as 7 kpc.  However, with
the strong external potential (SMHD2), the interarm features are
strong in the outer regions, but at radii of $\sim$5 - 11 kpc they are
weak.  In the inner regions, at $R$ $\la$ 5 kpc, the interarm features
are again much stronger than the background; there are no strong
interarm features in the innermost region of SMHD1.  The reason for
this difference in interarm features is clear in the structure of the
arms themselves.  In the $F$ = 3\% model, the arms inside 7 kpc are
smooth.  In the $F$ = 10\% models, on the other hand, the arms are
broken into many clumps.  The strong external potential in model SMHD2
has gathered more matter into the spiral arms, and self-gravity causes
concentrations to grow, with much of the gas eventually collapsing
into clumps.  Gas flowing through the arms can be concentrated by
these growing clumps, and returned to the interarm regions as
overdense spurs.  Since the stronger spiral potential of model SMHD2
concentrates more gas in the arms, the interarm regions at radii of
$\sim$5 - 11 kpc has less gas compared to model SMHD1 with a weak
spiral potential; as a result the interarm features in this region are
weaker in model SMHD2.

The boxed region from Figure \ref{SMHD1_2inpl}(b) is shown in detail
in Figure \ref{SLvec} with the instantaneous velocities, including the
unperturbed velocity field.  Far from the spiral arms, the
instantaneous velocities do not differ much from the initial circular
velocities.  As expected, the velocities of gas near the arms is
significantly perturbed.  Further, the over-dense clumps in the arm
flow along the arm.  At this stage, gravitationally bound structures
do not leave the gaseous spiral arm, but continue to build in mass as
matter from the interarm regions flows into the arm.  If kept
unchecked, the arm clumps would grow in a runaway fashion.  

The overlaid contours in Figure \ref{SLvec} indicate magnetic field
lines.  Initially, the magnetic field is directed only in the
azimuthal direction.  As the spiral arms increase in density, the
magnetic field is concentrated in the arms, thereby weakening the
field in the interarms.  The growth of the clumps along the arms
further perturbs the field lines.  However, only strong density
enhancements produce field perturbations; the interarm features
emerging from the clumps and the background features that grow in
models without an external spiral potential do not strongly affect the
magnetic field.

Figure \ref{SHD1} shows snapshots from a model without magnetic
fields.  Both the interarm features and clumps within the arms grow
much more rapidly compared with the corresponding model with magnetic
fields, SMHD2 (in Fig. \ref {SMHD1_2pl}).  As early as $t/t_{orb}$ =
1.0, the arm coherence in the outer regions is weakened due to the
rapid growth of over-dense regions.  Even in models with $\beta=10$
(not shown here), the arms rapidly fragment.  Similar to the models
without self-gravity, strong magnetic fields act to preserve the
overall arm shape, and suppress the growth of interarm features,
either those caused by self-gravity, or by hydrodynamic effects.

Interarm features also grow more rapidly as the initial Toomre
parameter $Q_0$ is reduced.  Figure \ref{SMHD3} shows the snapshots of
SMHD3, a model similar to model SMHD2 except that $Q_0=1$ instead of
2.  Even though this model has magnetic fields, the interarm features
still grow relatively rapidly.  Similar to low $Q$ models without an
external potential, this model is more susceptible to the growth of
perturbations at smaller radii than the $Q_0$ = 2 model.

We have also explored models with varying values of the pattern speed
and spiral pitch angle.  Figure \ref{SMHD4_5}(a) shows snapshots after
1 orbit of model SMHD4.  Model SMHD4 is similar to the fiducial model
SMHD2, except that $\Omega_p$ = 42 \kms~kpc$^{-1}$.  The corotation
radius for such a pattern speed is 5 kpc.  Upon first glance, this
snapshot may seem rather similar to the snapshot of model SMHD2 in
Figure \ref{SMHD1_2pl}.  However, in regions outside corotation the
interarm features protrude inwards, towards the galactic center.  The
arms still have the over-dense knots, but the stretched features near
the arms project in the opposite direction from those in the
fiducial model.  This is as expected, because outside corotation, the
rotating spiral potential has a greater angular velocity than the gas.
The flow enters the arms from the outer (convex) side, and leaves from
the inner (concave) side.  Azimuthally varying over/under dense
regions that are created within the arm and return to the interarm
region on the inside of the arm are sheared into trailing structures.
This reversed orientation is even more apparent in models where the
Toomre parameter $ Q \propto \, R$, presented in $\S$\ref{incrQmods}.
The trailing features in the outermost part of the disk arise in a
different way, however, as we shall discuss in $\S$\ref{disc}.

Model SMHD5, shown at $t/t_{orb}$ = 1 in Figure \ref{SMHD4_5}(b), is
similar to the fiducial model but with a larger pitch angle of $i$ =
20\degr.  Besides the expected difference in shape of the spiral arms,
many of the other features evident in SMHD2 (in Figs.
\ref{SMHD1_2pl}(c) and (d)), such as the knots of matter in the arm
and the trailing features in the outermost and innermost regions, are
also present.

\subsection{Spiral Models with $Q \propto R$ \label{incrQmods}}

For spiral models with initial surface density distributions $\propto
R^{-2}$, we only present cases with magnetic fields ($\beta = 1$); we
have shown that magnetic fields act to keep the arms intact.
Otherwise, self-gravity causes the runaway growth of the clumps in the
arms.  Figure \ref{rsq} shows snapshots of model SMHD3$^\prime$ ($Q_0$
= 1) and SMHD2$^\prime$ ($Q_0 = 2$).  Comparing with Figures
\ref{SMHD1_2pl}, \ref{SMHD1_2inpl}, and \ref{SMHD3}, these models are
much more stable in the outer regions, as expected.  Nevertheless,
strong interarm trailing structures do grow in the inner (14 kpc)$^2$.
Since these models are more stable, the interarm features do not
extend as for away from the spiral arms as those in the constant $Q$
models.  These features grow in both $Q_0=1$ and $Q_0=2$ models, but
are stronger in the $Q_0=2$ case.  It is also clear that the interarm
features connect with the most dense clumps in the arms, which are
more dense in model SMHD2$^\prime$ than SMHD3$^\prime$.  

Even though the self-gravitational force is stronger in model
SMHD3$^\prime$ ($Q_0=1$) than in model SMHD2$^\prime$ (because the
absolute $\Sigma$ is larger in SMHD3$^\prime$), the clumps in the arms
of model SMHD2$^\prime$ are more dense (relative to $\Sigma_0$).  This
results because more gas flows into the spiral arms in the more stable
disk of model SMHD2$^\prime$.  As shown from the corresponding disk
stability tests in Figure \ref{Qvartest}, the inner regions of models
with $Q_0=1$ are much more unstable than models with $Q_0=2$.  These
background instabilities will grow regardless of the presence of an
external spiral potential.  For a stable disk as in model
SMHD2$^\prime$, stability of the background disk allows more gas to
flow into the arms, resulting in stronger arms as well as clumps.

Figure \ref{rsqRcr5} shows a snapshot of model SMHD4$^\prime$, with
large pattern speed ($\Omega_p$ = 42 km s$^{-1}$ kpc$^{-1}$) at
$t/t_{orb}$ = 1.25.  The corotation radius of 5 kpc is indicated as
well.  For this model, $Q$ in the initial conditions varies from 0.4
at the inner radius to 6.3 at the outer radius.  The nature of the
interarm features in this model is much more clear than in the
corresponding model with constant $Q$ (SMHD4, shown in Fig.
\ref{SMHD4_5}(a)).  Inside corotation, the interarm features, which
are connected with the arm clumps, occur exterior to the main spiral
arms.  However, outside corotation, the interarm sheared features
emanate inwards from the main arms in the opposite sense from those
inside corotation.  This direction is downstream from the arms, as
seen in a frame rotating at $\Omega_p$.  Near corotation, clumps in
the arms exist, but do not extend much either interior or exterior to
the arms.

\section{Analysis and Discussion \label{disc}}
\subsection{Arm Spurs or Sheared Background Features? \label{asp}}
In our presentation and description of the models in $\S$\ref{sgsec},
we have referred to the interarm structures we identify as features,
not spurs.  We will define spurs as interarm features that are
distinctly associated with spiral arms, intersecting the main spiral
arms at locations where self-gravity caused the growth of clumps.
Using this definition, spurs in the present global models would
therefore be analogies of the structures studied in the local models
of KO.

Of course, it is well known that self-gravitating instabilities grow
when the surface density is large enough, whether or not there is an
external potential.  The resulting over-dense entities, which grow via
swing amplification, are stretched due to the shear in the disk.  The
interarm features arising in our models with spiral perturbations have
similar shapes to those in the stability test models, since the
interarm shear profile is similar to that of the unperturbed velocity
field.  At first glance, it is therefore not obvious whether the
interarm features in the spiral models are specifically due to the
spiral perturbation, or whether they would arise regardless of the
presence of the spiral perturbation.

Given, however, the dependence of the orientation of dense interarm
features on the spiral pattern speed (or the corotation radius) as
seen in Figures \ref{SMHD4_5}(a) and \ref{rsqRcr5}, it is clear the
external potential can have a significant effect.  Depending on
whether the spiral potential sweeps through the gas (outside
corotation), or whether the gas overtakes the spiral potential (inside
corotation), the interarm features in these cases grow inward or
outward from the arm, respectively.  This reversal of orientation
indicates that the growth of such features is dependent on the spiral
potential.  These dense features, furthermore, are all connected to
distinct arm clumps; they therefore fit our definition of ``true
spurs.''  On the other hand, the lower density interarm features
evident in Figures \ref{SMHD1_2pl}(b) and \ref{SMHD3}(b) are similar to
the structures seen in Figure \ref{nosp} that grow in the absence of a
spiral potential, provided the interarm surface density is
sufficiently high.  These ``background features'' are often {\it not}
associated with arm clumps when they are present; we therefore do not
consider them ``true spurs.''

We can quantify the effect of the spiral potential on outer-disk
features by computing the dimensionless wavenumber of the background
features $K_{y,feature} = \lambda_J/\lambda_{feature}$.  Here, $\lambda_J
= c_s^2/(G\Sigma)$ is the local two-dimensional Jeans length, and
$\lambda_{feature}$ is the mean azimuthal separation of the background
features.  Table \ref{wavenum} shows the feature separation and the
wavenumber for MHD models ($\beta=1$) with the external potential
strengths F = 0\%, 1\%, 3\%, and 10\%, for a region in the outer part
of the disk.  The other parameters are the same as the fiducial model:
$i$ = 10\degr, $\Omega_p$ = 8.4 \kms~kpc$^{-1}$, and $Q_0$ = 2 (with
initial surface density distribution $\propto R^{-1}$).  For the
feature separation $\lambda_{feature}$, we use the mean of the
distances between the peaks of the interarm features at a radius of
$R$ = 9.9 kpc, along an arc of 80\degr, at $t/t_{orb} = 1$ (the
feature separation does not vary much with time).  At $R$ = 9.9 kpc,
using the initial surface density we find $\lambda_J = 1.4$ kpc.  The
table shows that the feature separation does not vary much with the
strength of the external potential, suggesting that these outer-disk
features are not ``true spurs.''  In fact, the feature separation
is always approximately the Jeans length.  For model SMHD3 ($Q_0=1$,
$F=10\%$), the feature separation at the same radius is 1.2 kpc.  The
value of $\lambda_J$ is half that of the $Q_0=2$ model, while the
feature spacing decreases by 30\%.  This gives a ratio
$K_{y,feature}$ that is $\approx$ 30\% smaller when $Q=1$ than when
$Q=2$, but is still close to unity.

From Figure \ref{SMHD1_2inpl}, it is clear that a strong external
potential is required for interarm features to grow in the inner
regions.  Similarly, in $Q \propto R$ models (Fig. \ref{rsq}),
stretched interarm features only grow near the arms in models with a
strong external potential.  In models having $Q \propto R$ without an
externally imposed spiral potential (Fig. \ref{Qvartest}(b)), sheared
trailing features do not grow in the inner regions.  Further, the
features that do grow in models with strong spiral potentials are
connected to the clumps that form in the spiral arm itself.  Thus, we
identify the interarm features in the inner regions as true arm spurs,
and for measuring the separation we replace the symbol
$\lambda_{feature}$ by $\lambda_{spur}$.  At $R\approx$ 4.5 kpc, we
measure values of $\lambda_{spur}\approx$ 0.6 kpc, and
$\lambda_J\approx$ 0.6 kpc using the initial surface density.  We find
that the spur separation is again approximately equal to the Jeans
length at ambient densities.  If instead we had used the value of
$\Sigma$ in the arm, $\lambda_J$ would decrease, giving the ratio
$\lambda_{spur}/\lambda_{J,arm}\sim$ 5.\footnote{In this region,
  $\Delta R \sim 25$ pc and $R\Delta\theta \sim 15$ pc, so the Jeans
  length using arm densities is resolved.}  For model SMHD3 ($Q_0=1$),
both $\lambda_{spur}$ and the arm surface density (therefore
$\lambda_{J,arm}$) are comparable to those quantities in model SMHD2,
though the initial background surface densities differ by a factor of
two.  Thus, for $Q=1$ the ratio
$\lambda_{spur}/\lambda_{J,background}\approx2$, and $\lambda_{spur}/\lambda_{J,arm}\approx5$.  Evidently, for ``true spurs'' the spacing depends more
directly on $\lambda_{J,arm}$ than on $\lambda_{J,background}$.  Even
though spurs will grow in the inner regions only if there is a strong
spiral potential, the distance between the spurs is still within a
factor of two of the minimum scale length required for gravitational
instability under uniform conditions; when $Q=2$ the ratio
$\lambda_{spur}/\lambda_{J,background}$ is indistinguishable from the
case of ``background features'' that grow in the outer regions
independent of the spiral potential.  For realistic $Q$ values near 2,
the separation of filamentary structures is thus not sufficient in
itself to determine their origin.  The additional consideration of
whether structures are connected to arm clumps or not discriminate
between ``true spurs'' and swing-amplified ``background features.''

\subsection{Spurs and Arm Clumps \label{clumps}}
As discussed above, we term interarm features that grow out of arm
condensations ``true spurs.''  When interarm features grow as
background effects, strong condensations can also grow within and
remain in the arm.  In this case, however, there is not a one-to-one
relationship between interarm features and arm condensations (see e.g.
the outer regions of Fig.  \ref{SMHD1_2pl}(d) and Fig.
\ref{SMHD3}(a)).  In both situations, however, the arm condensations
that grow are generally regularly spaced, similar to the ``beads on a
string'' of bright HII regions often observed in spiral galaxies.  The
spacings between the clumps in the $Q_0=2$ model (SMHD2) at
$R\approx4.5$ kpc is typically $\sim$630 pc, comparable to
5$\lambda_{J,arm}$ (measured using arm surface densities before
fragmentation).  In other models with distinct arm clumps, such as
SMHD3 ($Q_0=1$), SMHD4 ($R_{CR}=5$ kpc) and SMHD5 ($i=20$\degr) we
measure clump spacings of $\sim$3 - 5$\lambda_{J,arm}$.

We also measured the width $W$ of the spiral arms.  For the fiducial
model SMHD2 ($Q_0=2$ and $F$ = 10\%), the FWHM $W\approx210$ pc, for
the same region of the arm for which the clump/spur spacing was
measured.  The ratio $\lambda_{spur}/W\sim3$ is consistent with the
observational study of Elmegreen \& Elmegreen (1983) and the
theoretical analysis of Elmegreen (1994).  For the $Q_0=1$ model,
SMHD3, we measure an arm width $W\approx600$ pc, and clump/spur
spacing of $\approx600$ pc.  Thus, when $Q_0=1$ the ratio
$\lambda_{spur}/W$ is close to unity.  The measured ratio
$\lambda_{spur}/\lambda_{J,arm}$ is therefore more consistent between
our differing $Q$ models than the ratio $\lambda_{spur}/W$, possibly
due to magnetic fields and the physics of MJI (see KO).  In practice,
however, the observed range of $Q$ might not be large enough to
distinguish a difference.

For models shown in Figure \ref{SMHD1_2pl} (and \ref{SMHD1_2inpl}),
only model SMHD2, with a strong spiral potential, shows distinct
clumps in the arms.  In the case with a weaker potential, the density
transitions smoothly from the arm to the interarm features.  Using the
clump finding algorithm {\tt clumpfind} (Williams et al.\ 1994), we
consistently measure the clump masses to be $M_{cl}\approx10^7
M_\odot$ in the arms of model SMHD2.  In terms of the Jeans mass
\begin{equation}
M_J = \frac{c_s^4}{G^2 \Sigma}
\label{jeansm}
\end{equation}
$M_{cl}\approx10M_J$ using mean arm surface densities.  We measure
similar values for the other models with distinct arm clumps.  We find
that altering the contour levels for the clump finding algorithm does
not significantly change the total mass of clumps, but only increases
the number of clumps found, giving similar masses for the new clumps.
For models with weaker spiral potentials, strong clumps are not found
in the spiral arms, so we cannot define clumps in the arm as easily;
as discussed, the interarm features in models with weak potentials are
not true arm spurs.

In observations of many galaxies, especially the HST and Spitzer
images of M51, and other galaxies in the SINGS sample (Scoville \&
Rector 2001, and Kennicutt et al.\ 2003), the strong interarm features
indeed tend to intersect the brightest regions in CO along the main
dust lanes (La Vigne et al.\ 2006).  Vogel, Kulkarni, \& Scoville
(1988) found molecular complexes with masses of $10^7 - 10^8 M_\odot$
in M51, which they named Giant Molecular Associations (GMAs).  In
spiral galaxies for which the gaseous component is not predominantly
molecular, large HI clouds have also been found to have masses of
$\sim$$10^7 M_\odot$; these are termed ``superclouds'' by Elmegreen \&
Elmegreen (1983).  Both the GMAs and superclouds are analogous to the
arm clumps in our simulations, which do not include the chemistry of
the gas.  In addition, La Vigne et al.\ (2006) measure feather
spacings of $\sim$7 - 11$\lambda_J$ in M51 and $\sim$1.5$\lambda_J$ in NGC
0628, using surface densities in the arm to compute $\lambda_J$.
These measurements assume the same value of $c_s$ for both cases, and
may be affected by uncertainty in the conversion of CO luminosity to
gas mass.  The consistency of clump masses and spur separations and
the clump/spur connection in our models to the GMA masses, interarm
feather separations, and GMA/feather association in M51 suggests that
the strong spiral potential is directly responsible for producing
these structures.

The orientation of spiral arm spurs indicates whether a given region
is inside or outside corotation.  Thus, the location of corotation can
be identified if the transition from inward to outward directed spurs
is observed.  As discussed in KO, however, there are presently no
known galaxies that exhibit clear inward projected spurs for a number
of possible reasons, such as relatively weaker arms outside
corotation, and current resolution limits.

\subsection{Offset between Gaseous Arm and Potential Minimum \label{gccomp}}

As indicated in $\S$\ref{nsg}, the relative location between the
gaseous arm peak and the minimum in the spiral potential varies
depending on the strength of the potential and the corotation radius.
Figure \ref{offset} shows the azimuthal locations of these peaks,
which would be observed as the main dust lanes, for three models
relative to the potential minimum, as a function of radius.  The
location of the potential minimum for models HD1 and HD2 at any given
time is the same; only the strength of the potential differs.  Inside
corotation, the gaseous arms from models (including those not shown
here) with stronger potentials form closer to, though always
downstream from, the potential minimum.  As shown in Figure
\ref{offset}(b), the gaseous arm shifts from downstream to upstream
from the potential minimum at corotation.

Gittins and Clarke (2004), hereafter GC, find via the one-dimensional
shock-fitting procedure of Shu, Milione, \& Roberts (1973) with local
non-self-gravitating models, that the gas shock occurs upstream from
the potential minimum.  The magnitude of the offset depends on various
parameters.  They find that this offset approaches $-\pi$ at
corotation, suggesting that the location of corotation can be
constrained by measuring offsets between the arm in K band (tracing
the potential) and molecular (gas) observations.  In the cases studied
by GC, the potential minima and the gaseous shock intersect well
inside corotation.  The location of this intersection varies depending
on the strength of the spiral potential.

There are a number of possible reasons for the differences in offsets
between our models and the results of GC.  Perhaps most importantly,
we use a flat rotation curve, whereas GC uses a velocity profile that
varies with radius.  In addition, we measure the position of the
density peak (shocks are difficult to distinguish near corotation at
our resolution) while GC report on the position of the shock, which
may be upstream from the density peak.  Such differences between the
parameters and analysis methods in our models and those of GC prevent
a direct comparison of the results.  We note that both studies include
the effect of the stellar disk only as a fixed rotating spiral
potential, i.e.\ an unresponsive component.  The relative locations of
the gaseous and stellar arms may well depend on the mutual
self-consistent interaction between the two components, an important
issue for future investigation.  

\subsection{Effect of Disk Thickness on Stability \label{Qeff}}
In our calculation of the gaseous self-gravity, we (approximately)
include the effect of disk thickness $H$ (see Appendix).  Including
this disk thickness approximation, the local axisymmetric dispersion
relation for an unmagnetized medium becomes:
\begin{equation}
\omega^2 = \kappa^2 +c_s^2 k^2 - \frac{2\pi G \Sigma|k|}{1+|k|H}.
\label{effdisp}
\end{equation}
When $H=0$, the dispersion relation takes on its familiar form for
razor thin disks (e.g. Binney \& Tremaine 1987).  In order to solve
for the minimum value of $\omega$, we define
\begin{equation}
k_0 \equiv \frac{\pi G \Sigma}{c_s^2} = \frac{\kappa}{Q c_s},
\label{ko}
\end{equation}
and
\begin{equation}
y \equiv k/k_0,
\label{yk}
\end{equation}
so that at $\omega_{min}$,
\begin{equation}
y(1+k_0Hy)^2 = 1.
\label{yeqn}
\end{equation}
The critical value of $Q$ (where $\omega_{min}^2$ = 0) is then given by
\begin{equation}
Q_{crit}^2 + y^2 -\frac{2yQ_{crit}}{Q_{crit} + y(\frac{\kappa H}{c_s})} = 0.
\label{Qcrit}
\end{equation}

We solve equations (\ref{yeqn}) and (\ref{Qcrit}) simultaneously, and
Figure \ref{Qcritfig} shows how $Q/Q_{crit}$ varies with $R$, for
disks with $Q$ = 2.  With the modified dispersion relation of equation
(\ref{effdisp}), and $H$ = constant, disk stability decreases with
increasing $R$.  As shown in Figure \ref{Qcritfig}, in the inner
regions disks with large values of $H$ are much more stable than razor
thin disks, for which $Q_{crit}=1$ (Toomre 1964).

The stability profile for axisymmetric perturbations will also
influence the stability of non-axisymmetric perturbations (e.g.
Goldreich \& Lynden-Bell 1965, Toomre 1981, and Kim \& Ostriker 2001).
As seen in Figure \ref{nosp}, the outer regions of constant $Q$,
constant $H$ disks are unstable.  The maximum radius of the inner
stable region, $R_{crit}$, depends on $Q$, and is constant in time.
Models with varying thicknesses show that $R_{crit}$ increases with
larger values of $H$, qualitatively consistent with the stability
profiles in Figure \ref{Qcritfig}.  We note that $R_{crit}$ in our
models with different values of $H$ does not simply correspond to
$Q/Q_{crit}$ = constant, however.  For example, with $H$ = 25 pc, we
measure $R_{crit}$ = 5.2 kpc empirically, and $Q/Q_{crit}$ = 2.4 at
this location.  On the other hand, for $H$ = 200 pc, we measure
$R_{crit}$ = 7.8 pc, and $Q/Q_{crit}$ = 3.7 at this location.  Thus,
the value of $Q$ obtained from a modified dispersion relation
(including the effect of thickness) is not sufficient for a complete
characterization of non-axisymmetric stability limits in global
models.

We also find that in models with smaller values of $H$, the
instabilities (outside $R_{crit}$) grow sooner than in models with
larger values of $H$.  Further, the spacing between the perturbations
also vary if $H$ is altered significantly.  We find that increasing
$H$ by a factor of 8 increases the perturbation separation by a factor
of $\sim$2.5.  Slightly varying the thickness does not strongly affect
the perturbation spacing, however.  We also note that in our
implementation $H$ = constant, which favors structures to grow in the
outer regions; other choices of the thickness profiles, such as $H/R$
= constant, can result in stable outer regions combined with unstable
inner regions.

\section{Summary \label{summ}}

We have investigated the growth of interarm features in
self-gravitating gaseous disks of spiral galaxies using global MHD
simulations.  Our models are two-dimensional, but we account for the
thickness of the disk in an approximate way in the computation of
self-gravity.  Gaseous spiral arms grow as a result of an
externally-imposed rotating spiral potential, representing the stellar
spiral arms of a galaxy.  We explore a range of values for the
physical parameters describing the properties of the disk.  The main
results are as follows:

1) In the inner regions of disks without self-gravity or magnetic
fields, we are able to reproduce the interarm features that WK found
in their models.  When spiral shocks are strong enough, hydrodynamic
instabilities cause the growth of knots in the spiral arms, and the
shear causes interarm features to spread from these knots.  However,
we find that inclusion of magnetic fields gives more tensile strength
to the spiral arms, and suppresses the growth of such features.

2) In disks with a low amplitude (external) spiral potential but
without self-gravity, our simulations show long lasting spiral
patterns in the gas.  We also obtain bifurcation features (arm
branches) near locations of Lindblad and ultraharmonic resonances,
similar to features discussed in CLS.

3) To assess the intrinsic stability of disks for growth of moderate
($\sim$kpc) scale structure, we simulated self-gravitating disks
without an external spiral potential.  Slightly over-dense regions
(nonlinearly) grow in density due to self-gravity, and subsequently
become stretched due to the rotational shear.  For disks with a
constant Toomre parameter $Q$ (equivalently, with initial surface
density distribution $\propto R^{-1}$) and thickness $H$, the
instability initially grows in the outermost regions.  Only the inner
regions remain stable.  For disks with a lower value of $Q_0$, the
size of the inner stable region decreases, and the instabilities grow
sooner.  As the thickness of the disk decreases, so does the size of the inner stable region.  

4) In self-gravitating disk models with an external spiral potential,
interarm feather-like features can arise in two distinct ways.  One
way these features can develop is essentially the same as in models
without spiral structure summarized in (3) above, i.e.\ via swing
amplification in the interarm region.  Stronger spiral potentials
changes the spacing of these features only slightly.  However, strong
spiral potentials can lead to growth of self-gravitating knots strung
out as beads along the arms.  Growing arm condensations in turn can
produce interarm feathering in a second way, by concentrating gas as
it flows from the arm into the interarm regions.  If we define ``true
spurs'' as distinct interarm features associated with the brightest
regions in the arm, these structures only form in self-gravitating
disks with strong ($F \approx$ 10\%) external potentials.

5) Bound clumps that grow in the spiral arms have masses of
$\sim$$10^7M_\odot$.  The spacings of these clumps, or equivalently,
spurs, are measured to be $\sim$3 - 5$\lambda_{J,arm}$.  In models
where $Q_0=2$, the ratio of clump spacing to arm width is consistent
with the prediction $\lambda/W\approx3$ from Elmegreen (1994) and the
observational study of Elmegreen \& Elmegreen (1983).  In many
galaxies large clouds, such as GMAs and superclouds, are observed to
have similar masses and spacings to the knots in our models.

6) We find that without magnetic fields, the arms in self-gravitating
models rapidly fragment, destroying the continuous, distinct spiral
arm shape.  Thus, magnetic fields may be important for maintaining the
integrity of grand design spiral structure in the ISM, even as
self-gravity (and star formation) work to destroy these large-scale
patterns.

7) The distribution and extent of interarm features that grow in
self-gravitating models depend on the surface density distribution
(or, equivalently, the instability parameter $Q$).  For $R^{-1}$
surface density distributions, for which $Q$ is initially constant,
strong feathering grows early on in the outer regions of the disk.  At
later times in the inner regions, spurs grow from the arm clumps and
extend out to nearly the adjacent arm.  For $R^{-2}$ surface density
distributions, $Q$ increases with radius.  For these disks, spur
formation in the outer regions is suppressed, but prominent spurs
still grow in the inner regions.  Our models have adopted $H$ =
constant; if $H$ increases with $R$, it is possible to have inner-disk
instability and outer-disk stability even with $Q$ = constant.

8) The orientation of the spurs with respect to the arms depends on
the pattern speed of the spiral potential.  Inside corotation, spurs
extend outward from the convex side of the main dust lanes, as this is
the downstream side of the arm.  Outside corotation, the potential
rotates faster than the gas, so the spurs form inside the main dust
lanes.  In principle, a reversal in the orientation of spurs in
observed galaxies could be used to determine the position of the
corotation radius.

Though we were able to produce spurs and quantify the conditions
necessary for the growth of spurs with the models of this work, we
were unable to follow the subsequent evolution of the disk for many
orbits.  Runaway growth of the massive clumps in the spiral arms
causes the surrounding gas to have large velocities, which implies
short timesteps in order to satisfy the Courant condition.  In
addition, the densities become sufficiently large and the clumps so
small that they are not well resolved by our grid.  However, the
physicality of the simulation itself can be questioned at late times,
because real condensations would not grow uninhibited.  The
agglomeration of gas in galaxies into GMCs eventually leads to the
formation of stars.  The photoionization and mechanical inputs from
HII regions and supernovae associated with massive star formation in
GMCs returns much of their gas to the diffuse phase.  In future work,
we plan to implement feedback mechanisms, using appropriate energy
injection rates from typical star formation processes to disperse the
clumps formed in the arms.  We also plan to include heating and
cooling processes to simulate a realistic multi-phase medium.  By
including processes of this kind, it will be possible to study the
spur structure morphology and GMC formation rates and properties
consistent with quasi-steady-state conditions.

\acknowledgments

Acknowledgments: We thank W. T. Kim and S. N. Vogel for stimulating
discussions, and G.  C. Gomez for his role in implementing cylindrical
polar coordinates in ZEUS.  We are also grateful to P. J. Teuben for
assistance using {\tt NEMO} software (Teuben, 1995) for data analysis
and visualizations.  We thank B. G. Elmegreen for a helpful and
insightful referee's report.  This research is supported by the
National Science Foundation under grants AST-0228974 and AST-0507315.

\appendix
\section{Appendix}
\subsection{Self-Gravity: Cartesian Coordinates \label{cartesian}}
Numerically, Poisson's Equation
\begin{equation}
{\nabla^2 \Phi} = 4\pi G\rho,
\label{poisson}
\end{equation}
can be solved efficiently using Fourier methods.  In 3D Cartesian
coordinates, 
\begin{equation}
{\hat{\Phi} (k_x,k_y,k_z)} = -\frac{4\pi G\hat{\rho}(k_x,k_y,k_z)}{k^2},
\label{3D_cart}
\end{equation}
where $\hat{\Phi}$ and $\hat{\rho}$ are the Fourier transform of the
potential $\Phi$ and density $\rho$, and $k^2 = k_x^2 + k_y^2 +
k_z^2$.  With density periodic on a domain $(L_x,L_y,L_z)$ with
$(N_x,N_y,N_z)$ zones in each dimension, respectively, the values of
$k_x = \pm n_x 2 \pi / L_x$ with $n_x = 1\, .\,.\,.\, N_x/2$.

For an infinitesimally thin, two dimensional disk, one may use separation of variables to show that the potential within the disk in Cartesian coordinates satisfies 
\begin{equation}
{\hat{\Phi} (k_x,k_y)} = -\frac{2\pi G\hat{\Sigma}(k_x,k_y)}{\left| {\bf k}\right|},
\label{2D_cart}
\end{equation}
where $\hat{\Sigma}$ is the Fourier transform of the surface density
$\Sigma$ (e.g. Binney \& Tremaine 1987) .  The effect of the nonzero
disk thickness may be accounted for in an approximate way (Kim,
Ostriker, \& Stone 2002), such that equation (\ref{2D_cart}) becomes
\begin{equation}
{\hat{\Phi} (k_x,k_y)} = -\frac{2\pi G\hat{\Sigma}}{| {\bf k}|(1+|{\bf k}|H)},
\label{2D_thick}
\end{equation}
where $H$ is the thickness of the disk, and $\Sigma = 2H\rho$.

Obtaining the potential using Fourier methods expressed in equations
(\ref{3D_cart}) or (\ref{2D_cart}) assumes that the density is
periodic.  Thus, the resulting potential includes a
contribution from replicas of the density distribution (outside the
computational domain).  If in fact one desires to compute the
potential of a spatially isolated system with zero density outside of
the computational domain, then this method is modified by computing
Fourier transforms on a larger, zero-padded array.  The central
portion of the larger array is filled with the density values from the
original domain, and the surrounding zones are set to zero.  For
obtaining a solution of the Poisson's equation the zero-padded array
must be at least twice the size of the original array in each
dimension (e.g. Hockney \& Eastwood 1981).  A larger padded region moves
unphysical cusps away from the boundaries of the computational domain;
of course, increasing the size of the padded region requires more
memory as well as CPU time.

\subsection{Self-Gravity: Polar Coordinates \label{cyl}}

In 2D cylindrical polar coordinates, a number of techniques have been
explored to calculate the potential.  Miller (1978) describes a method
which sums the potential due to concentric rings in polar coordinates.
The potential at $(R,\phi)$ is written as
\begin{equation}
\Phi(R,\phi) = -G \int_{0}^{\infty} R^{'}dR^{'} \int_{0}^{2\pi}d\phi^{'} D(R,R^{'};\phi-\phi^{'})\Sigma(R^{'},\phi^{'})
\label{Miller_phi}
\end{equation}
where 
\begin{equation}
D(R,R^{'};\phi-\phi^{'}) = {(R^2+R^{'2}+\epsilon^2-2RR^{'}\cos(\phi-\phi^{'}))^{-\frac{1}{2}}}.
\label{Miller_D}
\end{equation}
Here, $\epsilon$ is a softening parameter.  We can discretize equation
(\ref{Miller_phi}), writing the result as
\begin{equation}
\Phi(R_i,\phi_l) = -G\sum_{n} \sum_{j} D(R_i,R_j;\phi_l-\phi_n) \Sigma(R_j,\phi_n)\delta A_{jn}
\label{Miller_discrete}
\end{equation}
where $\delta A_{jn} = R_j\delta R_j \delta \phi_n$.  The direct
summation in equation (\ref{Miller_discrete}) is computationally
expensive; however, Fourier transforms can accelerate the computation.
The integral over $\phi^{'}$ in equation (\ref{Miller_phi}) is a
convolution of $D$ with the mass distribution in the ring at $R_j^{'}$
with the equivalent for the sum in equation (\ref{Miller_discrete}).
By the Fourier convolution theorem, we can write
\begin{equation}
\hat{\Phi}_n(R_i) = -G\sum_{R_j} \hat{D}_n(R_i,R_j)\hat{M}_n(R_j),
\label{Miller_FT}
\end{equation}
where $\hat{M_n}(R_j)$ is the 1D Fourier transform in the azimuthal
direction of the mass $M_{j,n} = \Sigma_{j,n} \delta A_{j,n}$.  The
discrete Fourier transform (in the $\hat{\phi}$ direction) of the
Green's Function $D(R,R^{'};\phi-\phi^{'}),\, \hat D_n(R_i,R_j),$ only
needs to be computed once, at the beginning of the simulation.  For an
$(R,\phi)$ computational grid, including periodic replicas of the
density is required to cover the full $2\pi$ domain of the azimuthal
coordinate.  The computational domain must therefore be $2\pi / m$ for
some $m$; we use $m=2$ for a two-armed spiral.  Aside from the effect
of softening, this method of computing the potential is exact.

Even though Miller's (1978) method is more efficient than direct
summation of equation (\ref{Miller_discrete}), more memory and CPU time
are still required than for a pure Fourier approach.  Thus, for
numerical expediency, we instead use a computationally more efficient
method to obtain an approximate solution for $\Phi$.  Though this
method is not an exact calculation of $\Phi$, we will show that the
differences between the exact (Miller) method and our more efficient
method is small.  After describing our method, we will show how the
results, as well as CPU usage, from the different methods compare.

To compute an approximate potential $\Phi$ in polar coordinates based
solely on Fourier transforms, we make use of the method described in
$\S$\ref{cartesian}.  We will hereafter refer to this method as the
coordinate transformation method.  If we apply a coordinate
transformation
\begin{eqnarray}
x & \equiv & R_0\ln\left(\frac{R}{R_0}\right) \nonumber \\
y & \equiv & R_0\phi,
\label{coordtrans}
\end{eqnarray}
Poisson's equation becomes
\begin{equation}
\frac{\partial^2\Phi}{\partial x^2} + \frac{\partial^2\Phi}{\partial
  y^2} + \left(\frac{R}{R_0}\right)^2\frac{\partial^2\Phi}{\partial z^2} = \left(\frac{R}{R_0}\right)^2 4 \pi G \rho.
\label{Poiss_trans}
\end{equation}
For in-plane gradients large compared to vertical gradients, the
solution is
\begin{equation}
\hat{\Phi}(k_x,k_y)= -\frac{2\pi G \hat{\tilde\Sigma}}{|k|^2H},
\label{FT_lipv}
\end{equation}
where $\tilde\Sigma = (R/R_0)^2\Sigma$.  This has the same form as
equation (\ref{2D_thick}) in the limit $|k|H >\!> 1$.  

More generally, solutions to the Laplace equation in cylindrical
coordinates have the form $\sum_{k,l}A_{kl}C_l(kR) e^{il\phi}
e^{-|kz|}$ where the $C_l$'s are two independent Bessel functions -
e.g.\ $J_l$ and $Y_l$ - and where $|z| \rightarrow \infty$ and
azimuthal boundary conditions have been applied.  The coefficients of
each term in the sum is determined by the requirement that
$\partial{\Phi} / \partial{z} |_{z=0^+} = 2\pi G \Sigma(R,\phi)$.
Each $A_{kl}$ can then be written as a Fourier-Hankel transform of the
surface density $A_{kl} \propto \int_{0}^{2\pi}e^{-il\phi}d\phi
\int_{-\infty}^{\infty}dR J_l(kR)R^2\Sigma(R,\phi)$ (e.g. Binney \&
Tremaine, 1987).

Since the force is dominated by terms with large gradients, the large
$k$ values are most important.  For large arguments, the Bessel
functions approach sinusoidal functions, so that the $A_{kl}$'s can be
written as two-dimensional Fourier transforms of $\tilde\Sigma$.  In
this limit, we must have $\hat\Phi(k_x,k_y)=-2\pi
G\hat{\tilde\Sigma}/|{\bf k}|$.  Altogether, we may therefore write our
approximate solution as
\begin{equation}
\hat{\Phi}(k_x,k_y)= -\frac{2\pi G \hat{\tilde\Sigma}}{|{\bf k}|(1+|{\bf k}|H)},
\label{FT_trans}
\end{equation}
Here, $k_y=l/R$ and $k_x$ is the wavenumber corresponding to the
transformed radial coordinate.  In order to have the $x_i$ values
equally spaced, the radial grid coordinates are equally separated in
$\log R$.

With this method, the surface density is implicitly assumed periodic
in both azimuth and radius.  The spurious contribution from periodic
replicas in the radial direction can be minimized by zero padding the
edges of the density array in the radial direction, as described at
the end of $\S$\ref{cartesian}.

\subsection{Comparison of Methods}

To compare the effects of the periodic replicas in the test cases for
the coordinate transformation method, we use arrays that are
2$\times$, 4$\times$, 8$\times$, 16$\times$, and 32$\times$ the size
of the original density array.  The computational domain for the
comparison tests has 256 radial elements and 1024 azimuthal elements.
As an example, for the case for which we use a 4$\times$ larger array,
the size of the $\tilde\Sigma$ array before taking a Fourier transform
to obtain $\hat{\tilde\Sigma}$ is 1024 $\times$ 1024.  The actual
values of the densities are stored in array elements where the first
(radial) index is between 385 and 640.  All other elements of the
array are set to zero.

To compare the two methods of computing the potential from test
simulations, we arrange test cases for which the total mass in the
computational domain is 0 or very small.  The Fourier method will
include contributions from matter outside the computational domain,
due to the assumed periodicity of the density.  Thus, minimizing the
total mass will reduce this superfluous contribution.  In analyzing
the differences between the methods, we keep in mind that the
softening will affect the numerical values of the potential (and
force) in regions with large density gradients.  With the coordinate
transformation method, we also compare the results from cases where
the size of the zero padded zones vary.

We inspect the potential of three test disks containing (1) a positive
and negative ring, (2) a positive and negative radial spoke, and (3) and
a positive and negative logarithmic spiral arm.  We compare the region
in between the positive and negative mass distributions regions, to
avoid edge effects.  For test cases (1) and (2), we find the relative
difference between the coordinate transformation and the Miller method
($|\Phi_{CT} - \Phi_{MM}|/\Phi_{MM}$) to be within $\sim$3\%.  For test case
(3), we measure a relative difference of $\sim$5\%.  Again, these
relative differences are measured between the positive and negative
density regions, away from the edges of the disk.  For the
self-gravitating simulations we perform, we are interested in the
growth of substructures under similar circumstances, i.e.\ away from
the edges of the disk, near regions where the perturbed density is
both positive and negative.  Nevertheless, even near the edges, all
tests give values of the potential that agree within $\sim$25\%.  Finally,
the difference between the tests using the coordinate transformation
(with varying sizes of the zero-padded zone) is negligible.

The main advantage of using the coordinate transformation method is
the decrease in CPU time for each simulation.  For this method
computing the potential requires a multiplication of the density to
obtain $\tilde\Sigma$, a Fourier transform to obtain
$\hat{\tilde\Sigma}$, a multiplication in Fourier space for the
gravitational kernel in equation \ref{FT_trans}, and an inverse
Fourier transform to obtain $\Phi$.  This sequence requires fewer
operations than the Miller computational method.
  
To measure the efficiency of each method, we use a slightly different
test simulation from the ones described above.  In these tests, with a
grid of 256 radial and 512 azimuthal zones, a spiral potential is
turned on over the first half orbit, then the self-gravitational
potential is slowly turned on over another half an orbit.  Thus, both
potentials are turned on fully after 1 orbit.  The test simulations
are subsequently allowed to run for an additional orbit, after which
the CPU times are compared.

All these tests, as well as many of the simulations presented in this
paper, were run on a machine with a 2.99 GHz Pentium 4 processor, with
2 GB of RAM.  Table \ref{CPUtab} shows the CPU time (relative to that
using the exact solution) required for each test to run.  It is
evident, as expected, that the simulations using the coordinate
transformation method (where we enlarge the density array by
2$\times$, 4$\times$, or 8$\times$) requires much less CPU time than
those using the exact potential computational method.  Since the
numerical differences between the methods are modest, and to take
advantage of the superior efficiency of the coordinate transformation
method, we use the coordinate transformation method, enlarging the
density array by 4$\times$, for our high resolution simulations
presented in this paper.  We have also tested models using the Miller
method, and obtain essentially the same results.

\vfill
\eject

\begin{deluxetable}{ccccc} 
\tablewidth{0pt}
\tablecaption{Parameters for Models Without Gas Self-Gravity}
\tablehead{
\colhead{Model} & \colhead{$\beta$} & \colhead{F (\%)} &
\colhead{$\Omega_p$ (\kms~kpc$^{-1}$)} &  
\colhead{$i$ ($\degr$)} \\
\colhead{(1)} & \colhead{(2)} & \colhead{(3)} & \colhead{(4) } &
\colhead{(5)}
}
\startdata
HD1 & $\infty$  & 3 & 8.4 & 10 \\
HD2 & $\infty$  & 10 & 8.4 & 10 \\
HD3 & $\infty$  & 10 & 42 & 10 \\
HD4 & $\infty$  & 10 & 8.4 & 20 \\
MHD1 & 1  & 10 & 8.4 & 10 \\
\enddata
\label{models}
\end{deluxetable}

\begin{deluxetable}{cccccc} 
\tablewidth{0pt}
\tablecaption{Parameters for Models Including Gas Self-Gravity}
\tablehead{
\colhead{Model \tablenotemark{a}} & \colhead{$\beta$} & \colhead{F (\%)} &
\colhead{$\Omega_p$ (\kms~kpc$^{-1}$)} &  
\colhead{$i$ ($\degr$)} & \colhead{$Q_0$} \\
\colhead{(1)} & \colhead{(2)} & \colhead{(3)} & \colhead{(4) } &
\colhead{(5)} & \colhead{(6)}
}
\startdata
SHDne1 & $\infty$  & 0 & - & - & 1 \\
SHDne2 & $\infty$  & 0 & - & - & 2 \\
SMHDne & 1  & 0 & - & - & 1 \\
\tableline
SMHD1 & 1  & 3 & 8.4 & 10 & 2 \\
SMHD2 & 1  & 10 & 8.4 & 10 & 2 \\
SMHD3 & 1  & 10 & 8.4 & 10 & 1 \\
SHD1 & $\infty$  & 10 & 8.4 & 10 & 2 \\
SMHD4 & 1  & 10 & 42 & 10 & 2 \\
SMHD5 & 1  & 10 & 8.4 & 20 & 2 \\
\enddata
{\singlespace
\tablenotetext{a}{Models listed in the text with a prime (e.g. SHDne1$^\prime$) have $\Sigma \propto R^{-2}$; otherwise, $\Sigma \propto R^{-1}$}
}
\label{sgmodels}
\end{deluxetable}

\begin{deluxetable}{ccc} 
\tablewidth{0pt}
\tablecaption{Scale of Interarm Features}
\tablehead{
\colhead{$F$ \tablenotemark{a}} & \colhead{$\lambda_{feature}$ \tablenotemark{b}} & \colhead{$K_{y,feature}$ \tablenotemark{c}}
}
\startdata 
0 & 1.3 & 1.1  \\
1\% & 1.3 & 1.1 \\
3\% & 1.4 & 1.0  \\
10\% & 1.7 & 0.8 \\
\enddata
{\singlespace
\tablenotetext{a}{\footnotesize External~Potential~Strength}
\tablenotetext{b}{\footnotesize Feature separation (kpc)}
\tablenotetext{c}{\footnotesize $\lambda_J/\lambda_{feature}$; $\lambda_J = \, 1.4$ kpc}
}
\label{wavenum}
\end{deluxetable}

\begin{deluxetable}{cccccc} 
\tablewidth{0pt}
\tablecaption{CPU Time for Different Potential Computation Methods}
\tablehead{
\colhead{CT (2$\times$) \tablenotemark{a}} & \colhead{CT (4$\times$)
  \tablenotemark{a}} & \colhead{CT (8$\times$) \tablenotemark{a}} &
\colhead{CT (16$\times$) \tablenotemark{a}} & \colhead{CT (32$\times$)
  \tablenotemark{a}} & \colhead{MM  \tablenotemark{b}}
}
\startdata
0.59 & 0.62  & 0.70 & 0.89 & 1.25 & 1.0 
\enddata
{\singlespace
\tablenotetext{a}{CT: Coordinate Transformation, with zero-padded enlargement of density array as indicated in parentheses}
\tablenotetext{b}{MM: Miller's Method}
}

\label{CPUtab}
\end{deluxetable}

\vfill
\eject

\begin{figure}
\caption{Snapshots from non self-gravitating, unmagnetized model HD1 (weak external spiral potential,  $F$ = 3\%).  Surface density at (a) $t/t_{orb}=1$ , when the external potential is turned on fully, (b) $t/t_{orb}=2$, and (c) $t/t_{orb}=3$.  (d) Inner 6.8 $\times$ 6.8 kpc$^2$ box shown in (c).  Color scale for (a), (b), and (c) is shown above (a) and (b), in units of log$(\Sigma/\Sigma_0)$.  Color scale for (d) is shown adjacent to (d), in units of $\Sigma/\Sigma_0$.}
\label{HD1pl}
\end{figure}

\begin{figure}
\caption{Snapshots from non self-gravitating, unmagnetized model HD2 (strong external spiral potential,
  $F$ = 10\%).  Surface density at (a) $t/t_{orb}=1$ , when the
  external potential is turned on fully, (b) $t/t_{orb}=2$, and (c)
  $t/t_{orb}=3$.  (d) Inner 6.8 $\times$ 6.8 kpc$^2$ box shown in (c).
  Color scales are arranged in the same manner as in Figure
  \ref{HD1pl}.}
\label{HD2pl}
\end{figure}

\begin{figure}
\caption{Snapshots at $t/t_{orb} = 2$, from models HD3 (a) and HD4
  (b).  Parameters are the same as in model HD2 (Fig. \ref{HD2pl}),
  but with (a) $\Omega_p$ = 42 \kms~ kpc$^{-1}$, and (b) $i = 20\degr$.
  The dashed circle in (a) indicates the corotation radius of 5 kpc.
  Color scale shows $\log(\Sigma/\Sigma_0)$.}
\label{HD3_4pl}
\end{figure}

\begin{figure}
\epsscale{1.1}
\plottwo{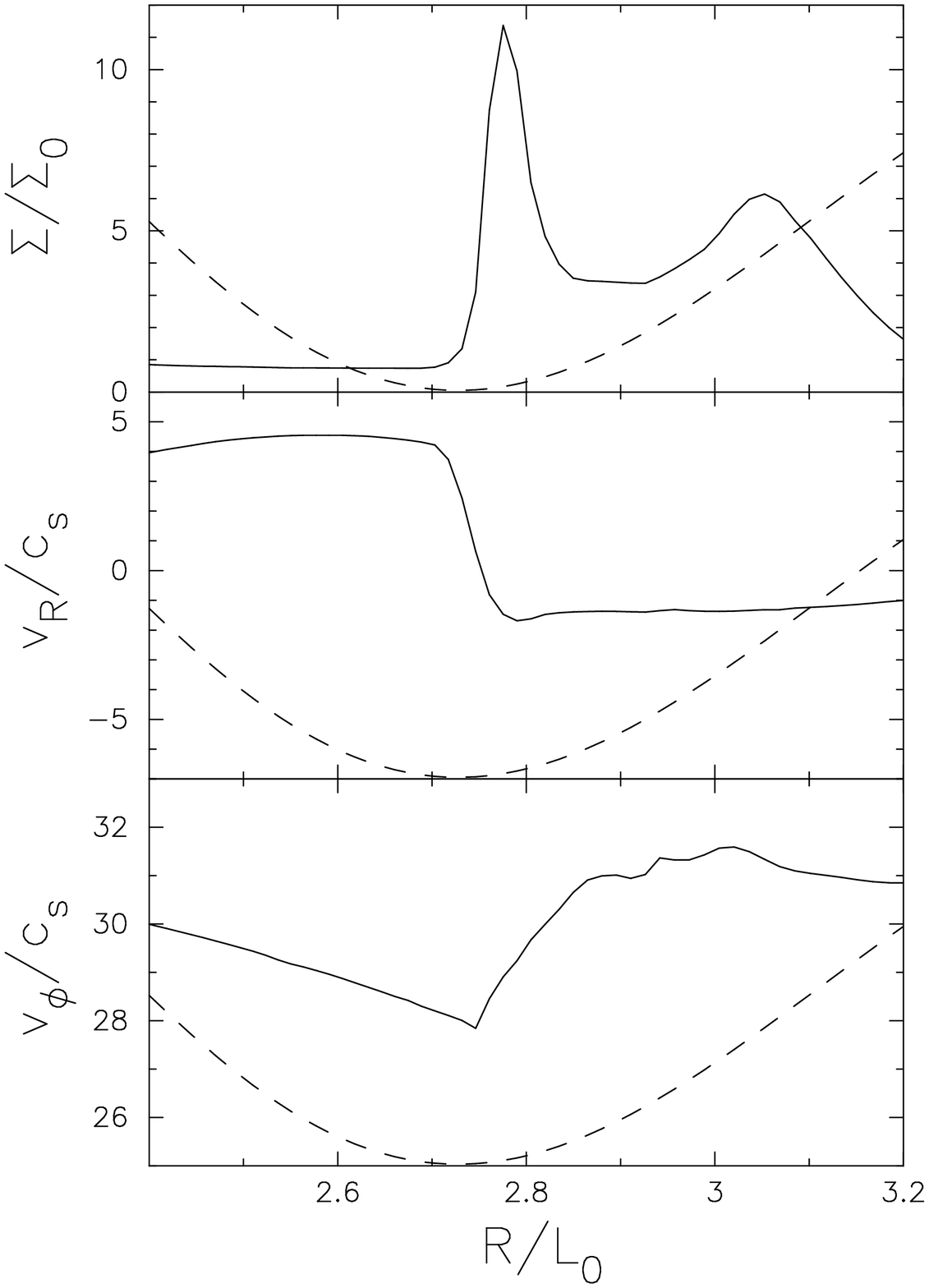}{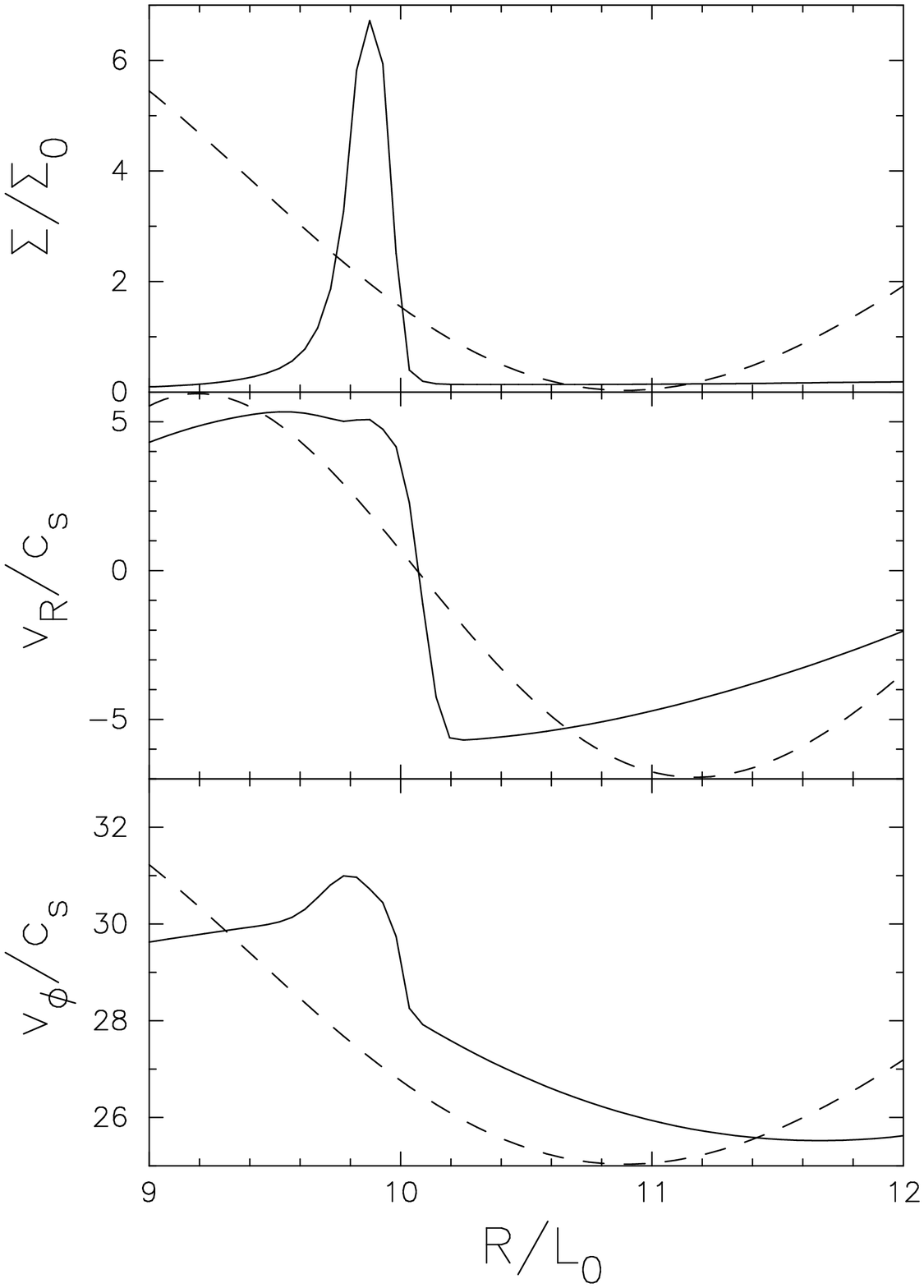}
\caption{Density, radial, and tangential velocity as a function of
  radius from model HD3 (shown in Fig. \ref{HD3_4pl}(a)), at time
  $t/t_{orb}$ = 1.26.  Quantities are taken from locations of constant
  azimuth, from a region inside corotation (left), and a region
  outside corotation (right).  The corotation radius is $R_{CR} / L_0$
  = 5.  The dashed line in each plot shows the external spiral
  potential.}
\label{profrel}
\end{figure}

\begin{figure}
\caption{Snapshots at $t/t_{orb} = 2$ of the inner 6.8 $\times$ 6.8 kpc$^2$ of model (a) HD2 and (b) MHD1.  Color scale is in units of $\Sigma/\Sigma_0$.}
\label{MHDcomp}
\end{figure}

\clearpage

\begin{figure}
\caption{Snapshots of models with self-gravity but no external potential.  (a) SHDne1
  ($Q_0 = 1$) at $t/t_{orb} = 0.97$, (b) SHDne2 ($Q_0 = 2$) at
  $t/t_{orb} = 0.97$, (c) SMHDne ($Q_0 = 1$, $\beta=1$) at $t/t_{orb}
  = 0.97$, and (d) SHDne2 ($Q_0 = 2$) at $t/t_{orb} = 1.49$.  Color
  scale is in units of $\log(\Sigma/\Sigma_0)$.}
\label{nosp}
\end{figure}

\begin{figure}
\caption{Snapshots at $t/t_{orb}$ = 1.0 of models with $Q \propto R$; (a) SHDne1$^\prime$ ($Q_0=1$) and (b) SHDne2$^\prime$ ($Q_0=2$).  Color scale shows $\log(\Sigma/\Sigma_0)$.}
\label{Qvartest}
\end{figure}

\begin{figure}
\caption{Models SMHD1 ($F=$ 3\%) and SMHD2 ($F=$ 10\%).  SMHD1 at
  (a) $t/t_{orb}$ = 1.0 and (b) $t/t_{orb}$ = 1.125.  SMHD2 at (c) $t/t_{orb}$ = 1 and (d)  $t/t_{orb}$ = 1.125.  The boxed regions in (b) and
  (d) are the inner 14 $\times$ 14 kpc$^2$ shown in detail in Figure \ref{SMHD1_2inpl}.  Units of color scale are $\log(\Sigma/\Sigma_0)$.}
\label{SMHD1_2pl}
\end{figure}

\begin{figure}
\caption{Inner 14 $\times$ 14 kpc$^2$ of Figure \ref{SMHD1_2pl}(b) and \ref{SMHD1_2pl}(d).  (a) Detail of $F$ = 3\% model from box shown in Figure \ref{SMHD1_2pl}(b), and (b) detail of $F$ = 10\% model from box shown in Figure \ref{SMHD1_2pl}(d).  Color scale shows $\Sigma/\Sigma_0$.  The boxed region in (b) is shown in Figure \ref{SLvec}.}
\label{SMHD1_2inpl}
\end{figure}

\begin{figure}
\caption{Boxed region from Figure \ref{SMHD1_2inpl}(b) (Model SMHD2).
  Solid vectors show the instantaneous gas velocity in the frame
  rotating with the spiral potential ($\Omega_p = 8.4$
  \kms~kpc$^{-1}$, $R_{CR}$ = 25 kpc).  Dotted vectors show the
  initial velocities (pure circular motion).  Scale of the vectors is
  shown by the thick vector (top right). Contours show magnetic field
  lines.}
\label{SLvec}
\end{figure}

\begin{figure}
\caption{Model SHD1 ($F=$ 10\%, $\beta=\infty$) at (a) $t/t_{orb}$ = 1.0, and (b) $t/t_{orb}$ = 1.125.  Color scale is in units of $\log(\Sigma/\Sigma_0)$.}
\label{SHD1}
\end{figure}

\begin{figure}
\caption{Model SMHD3 ($F=$ 10\%, $Q$=1) at (a) $t/t_{orb}$ = 0.75, and (b) $t/t_{orb}$ = 1.0.  Color scale shows $\log(\Sigma/\Sigma_0)$.}
\label{SMHD3}
\end{figure}

\begin{figure}
\caption{Snapshots at $t/t_{orb}$ = 1 of model (a) SMHD4 ($F$ = 10\%,
  $\Omega_p$ = 42 \kms~kpc$^{-1}$), with the corotation radius
  indicated by the dashed circle, and (b) SMHD5 ($F=$ 10\%, $i$ =
  20\degr ).  Units of color scale are $\log(\Sigma/\Sigma_0)$.}
\label{SMHD4_5}
\end{figure}

\begin{figure}
\caption{Snapshot of (a) SMHD3$^\prime$ ($Q_0=1$, $F=$ 10\%) at
  $t/t_{orb}$ = 0.875 and (c) SMHD2$^\prime$ ($Q_0=2$, $F=$ 10\%) at $t/t_{orb}$ = 1.125, along with the inner 14 $\times$ 14 kpc of each snapshot in (b) and (d).  Color scales of (a) and (c) are in units of $\log(\Sigma/\Sigma_0)$, and scales for (b) and (c) are shown in units of $\Sigma/\Sigma_0$.}
\label{rsq}
\end{figure}

\begin{figure}
\caption{Snapshot of model SMHD4$^\prime$ ($\Omega_p$ = 42
  \kms~kpc$^{-1}$) at $t/t_{orb}$ = 1.25. The corotation radius is
  indicated by the dashed circle.  Color scale is in units of
  $\Sigma/\Sigma_0$.}
\label{rsqRcr5}
\end{figure}

\begin{figure}
\includegraphics[angle=-90,scale=0.5]{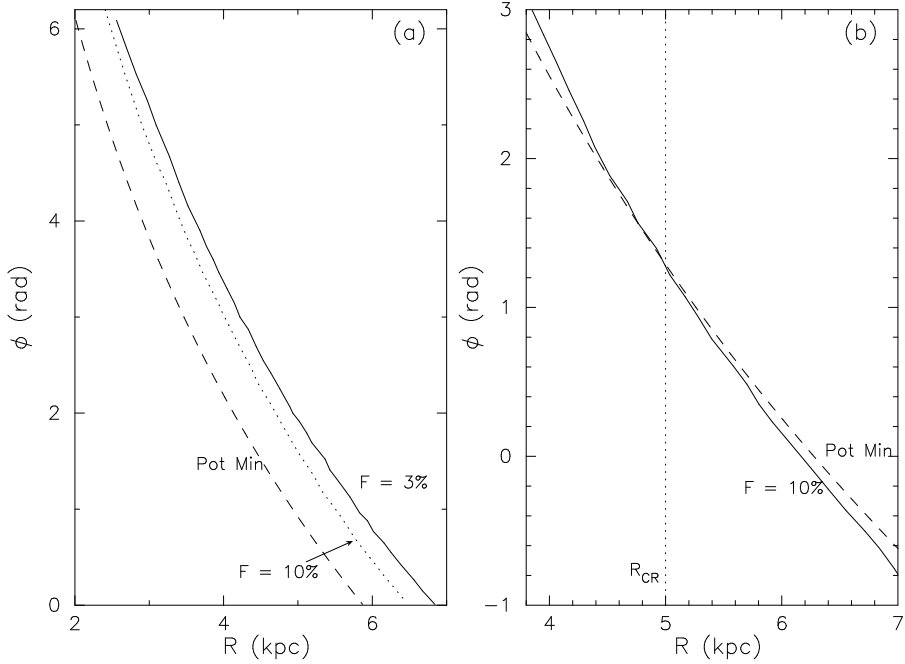}
\caption{Location of potential minimum and gaseous arm peaks.  (a) Models HD1 and HD2 with $F=$ 3\% and 10\%, respectively, and (b) model HD3, with $F$ = 10\% and $R_{CR}$ = 5 kpc.}
\label{offset}
\end{figure}

\begin{figure}
\includegraphics[angle=-90,scale=0.5]{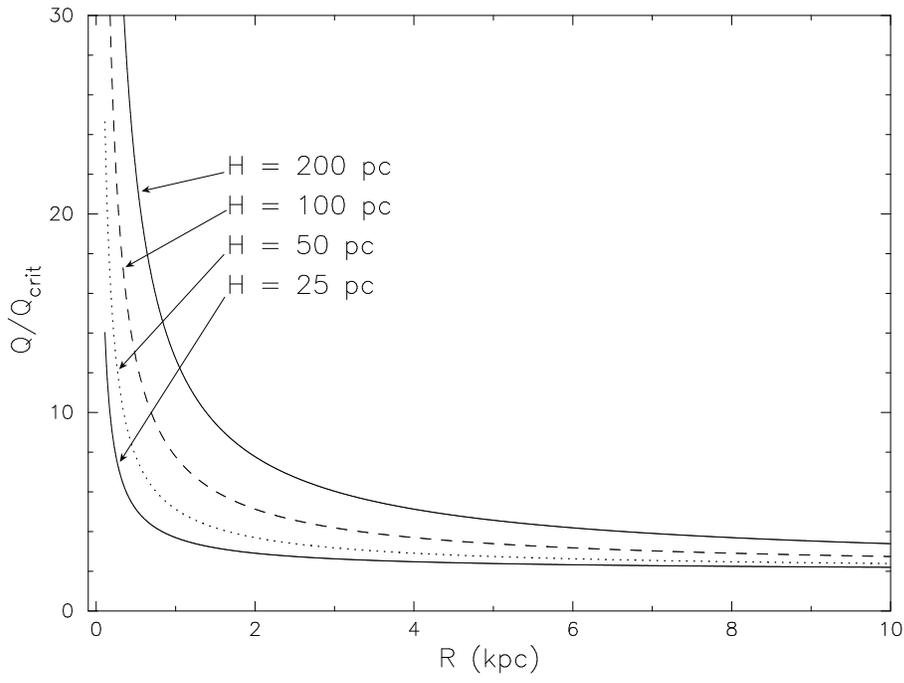}
\caption{$Q/Q_{crit}$ for $Q$ = 2 models when disk thickness is included (see eqns. [\ref{effdisp}] - [\ref{yeqn}]).}
\label{Qcritfig}
\end{figure}

\end{document}